# Coexistence of superconductivity and weak anti-localization at KTaO$_3$ (111) interfaces

Athby H. Al-Tawhid[1], Jesse Kanter[2], Mehdi Hatefipour[2], Divine P. Kumah[3], Javad Shabani[2], and Kaveh Ahadi[1,3,*]

[1]Department of Materials Science and Engineering, North Carolina State University, Raleigh, NC 27695, USA
[2]Center for Quantum Phenomena, Department of Physics, New York University, New York 10003, USA
[3]Department of Physics, North Carolina State University, Raleigh, North Carolina 27695, USA

[*] Corresponding author.  Email: kahadi@ncsu.edu

**ABSTRACT**

The intersection of two-dimensional superconductivity and topologically nontrivial states hosts a wide range of quantum phenomena, including Majorana fermions. Coexistence of topologically nontrivial states and superconductivity in a single material, however, remains elusive. Here, we report on the observation of two-dimensional superconductivity and weak anti-localization at the $TiO_x/KTaO_3(111)$ interfaces. A remnant, saturating resistance persists below the transition temperature as superconducting puddles fail to reach phase coherence. Signatures of weak anti-localization are observed below the superconducting transition, suggesting the coexistence of superconductivity and weak anti-localization. The superconducting interfaces show roughly one order of magnitude larger weak anti-localization correction, compared to non-superconducting interfaces, alluding to a relatively large coherence length in these interfaces.

A combination of broken inversion symmetry and strong spin-orbit coupling in two-dimensional (2D) superconductors gives rise to mixed-parity superconductivity [1], topological Weyl superconductivity [2], superconducting diode effect [3], and an upper critical field exceeding the Pauli-Chandrasekhar-Clogston limit [4,5]. Furthermore, the intersection of 2D superconductivity and topologically nontrivial states hosts a wide range of quantum phenomena, including non-abelian excitations [6], which are at the center of a groundbreaking proposal for fault-tolerant quantum computation [7]. 2D weak anti-localization has been used to probe surface states in topologically nontrivial systems [8,9]. The recent discovery of 2D superconductivity [10] and predictions of topologically nontrivial states [11] at the $KTaO_3$ (111) surface makes this material system a candidate platform for the coexistence of topologically nontrivial electronic states and unconventional superconductivity.

$KTaO_3$ is an incipient ferroelectric [12], in which superconductivity emerges at low temperatures in heavily doped samples [13]. A robust 2D electron system is reported at the interfaces of $KTaO_3$ with $LaTiO_3$ [14], $LaVO_3$ [15], $EuO$ [16], $LaAlO_3$ [17], $TiO_x$ [18] and $LaCrO_3$ [18]. The $KTaO_3$ conduction states are derived from Ta $5d$ and have a smaller effective mass and higher mobility and spin-orbit coupling compared to Ti $3d$ states in $SrTiO_3$ [19,20]. Spin-orbit coupling lifts the degeneracy of the Ta $5d$ states and splits them into $J=3/2$ and $J=1/2$ with a 0.4 eV energy gap, where $J$ is total angular momentum [21]. Recently, an exotic 2D superconductivity was discovered at the (111) [10] and (110) [22] $KTaO_3$ interfaces with $EuO$ and $LaAlO_3$, which shows near two orders of magnitude enhancement in the critical temperature of superconductivity ($T_C$) compared to its 3D counterpart [13]. Interestingly, $KTaO_3$ (100) interfaces do not show a superconducting transition. The superconducting state in the $KTaO_3$ (111) surface is highly susceptible to the interfacial structure, and a remnant resistance is observed below the superconducting transition temperature [23]. This failed superconductor state [24] is an ideal platform for the experimental realization of simultaneous superconductivity and nontrivial topology.

Here, we report on the observation of a superconducting transition at the $TiO_x/KTaO_3(111)$ interfaces. A true superconducting ground state ($R_s=0$), however, does not emerge as the superconducting puddles fail to reach phase coherence. Signatures of weak anti-localization are observed below the superconducting transition temperature, suggesting the coexistence of superconductivity and topologically nontrivial states at the $KTaO_3$ (111) surfaces.

Mobile carriers were introduced to the (111) surface of the KTaO3 single crystals using a 3 nm TiOx layer which induces oxygen vacancies. Here, the TiOx layer acts as an oxygen getter and is grown using an oxide molecular beam epitaxy system at a substrate temperature of 400 °C and $2\times10^{-10}$ Torr base pressure. The reflection high-energy electron diffraction (RHEED), measured during deposition, confirms the growth of an amorphous TiOx on the (111) KTaO3 surface (Supplementary materials, S1). Magneto-transport measurements were performed using the Van der Pauw configuration, and gold contacts were deposited using a sputter system at the corners of the samples through a shadow mask. The temperature-dependent magneto-transport measurements were carried out in a Quantum Design physical property measurement system (PPMS) with a lock-in amplifier (SR830, Stanford Research Systems) in AC mode with an excitation current of 10 µA and a frequency of 13.33 Hz. Sub-Kelvin magneto-transport measurements were carried out in a Triton dilution refrigerator, Oxford Instruments.

Oxygen vacancies introduce itinerant electrons to the Ta 5$d$ derived surface states. The conduction electrons at the low-temperature limit are derived from $J$=3/2, Ta 5$d$ states due to the large spin-orbit coupling gap in KTaO3 (0.4 eV [19,21]). Figure 1(a) shows a metallic behavior, d$R$/d$T$>0, in sheet resistance with temperature extending from room temperature to ~15 K. The sheet resistance changes somewhat linearly with temperature in this range. A resistance upturn emerges below 15 K, followed by a sharp drop below 3 K, Fig. 1(b). The abrupt drop in sheet resistance is consistent with recently discovered 2D superconductivity at the (111) KTaO3 interface [10]. Hall measurements were performed to determine the sheet carrier density. The Hall carrier density, $n = -1/(eR_H)$, where $R_H$ is the Hall coefficient and $e$ is the elementary charge. The Hall coefficient, $R_H = dR_{xy}/dB$, is extracted from a linear fit to the transverse resistance shown in Fig. 1(c). The sheet carrier density is ~$1\times10^{14}$ cm$^{-2}$ at 3 K, consistent with the peak of the critical temperature of superconductivity reported in (111) KTaO3 interfaces [10].

The residual resistivity ratio ($\rho_{300\ K}/\rho_{2\ K}$) is 2.3 and the carrier mobility increases from ~8 cm$^2$/Vs at room temperature to ~19 cm$^2$/Vs at 3 K. The moderate enhancement of the carrier mobility, despite the screening of the longitudinal optical phonons at low temperatures, can be explained by the interfacial scattering of itinerant electrons [25–27]. The spatial distribution of "two-dimensional" charge carriers controls their exposure to the interfacial structure and, as a result, the mean free path of charge carriers. Here, despite the modest low-temperature carrier mobility, the sheet resistance remains below the 2D Mott-Ioffe-Regel limit (~20 KΩ/□). Figure 1(b) shows a growing positive magnetoresistance with decreasing temperature (10-2 K). The positive magnetoresistance, particularly above 4 K, cannot be explained by the emergence of superconductivity alone and could be partially due to the weak anti-localization correction to the longitudinal resistance. 2D electron systems at the surface of $KTaO_3$ show large coherence length and signatures of weak anti-localization [18,28–30].

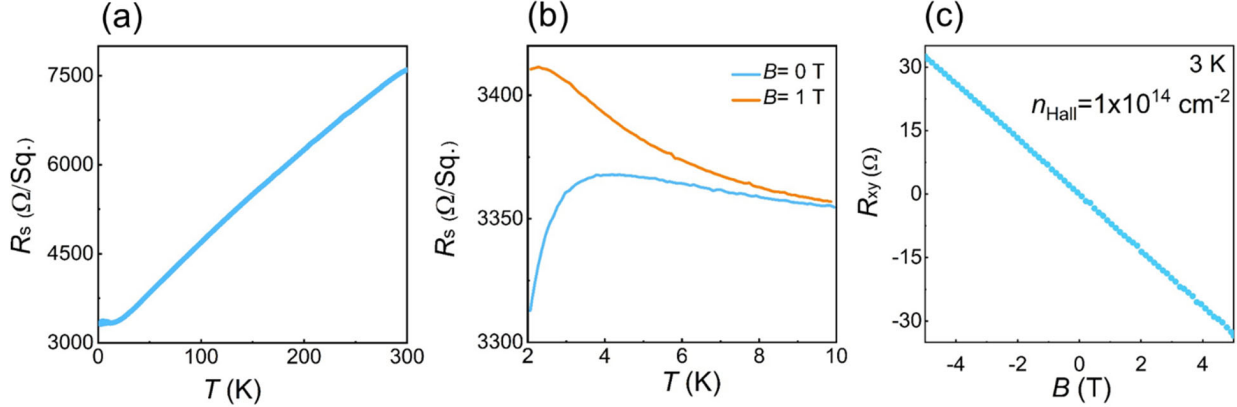

**Figure 1. Normal state electronic transport at the $TiO_x/KTaO_3$(111) interfaces.** (a) Sheet resistance with temperature (300-2 K) showing a linear scaling. (b) Magnetic field dependence of the sheet resistance-temperature behavior (10-2 K). (c) Transverse magnetoresistance at 3 K, resolving the 2D carrier density (~$1\times10^{14}$ cm$^{-2}$).

Figure 2(a) shows the normalized resistance with temperature from 20-0.1 K. The sharp drop in resistance is consistent with the observed superconducting transition at the interfaces of (111) $KTaO_3$ with EuO and $LAlO_3$ [10]. A remnant resistance, however, is observed below the superconducting transition temperature (mid-point $T_C$ ~1.1 K). The sheet resistance saturates to a nonzero value below the transition temperature, which is insensitive to the presence of filters, excluding the possibility of radiation thermalization. Furthermore, the $KTaO_3$(111)/$LaCrO_3$ interfaces, in which the normal state resistance is above the 2D Mott-Ioffe-Regel limit (~33 KΩ/□ at 3 K), do not show an abrupt drop in sheet resistance (Supplementary information, S2). Recently, a gate tunable remnant resistance was reported at the $KTaO_3$(111)/$LaAlO_3$ interfaces below the superconducting transition temperature [23], highlighting the role of interfacial structure on the emergence of a true superconducting ground state ($R_S$=0). A residual resistance has been observed in a wide range of 2D superconductors [31–34]. Here, the remnant resistance below the superconducting transition provides a unique platform for the experimental realization of 2D superconductivity coexisting with weak anti-localization.

The normalized longitudinal magnetoresistance, Fig. 2(b), shows that the relative change of resistance with the magnetic field ($R_{5T} - R_{0T}/R_{0T} = 0.24$, at 0.3 K) is large compared to the resistance change with temperature ($R_{3K} - R_{0.3K}/R_{0.3K} = 0.198$, at 0 T), alluding to the presence of both pair formation/breaking and weak anti-localization corrections in sheet resistance below the transition temperature. Furthermore, a sharp change in the resistance with the magnetic field is observed at low field (inset), consistent with the weak anti-localization [35]. The low field magneto-conduction, however, could not be explained by the Hikami-Larkin-Nagaoka model [36] due to the mixed weak anti-localization and superconducting corrections (Supplementary materials, S3).

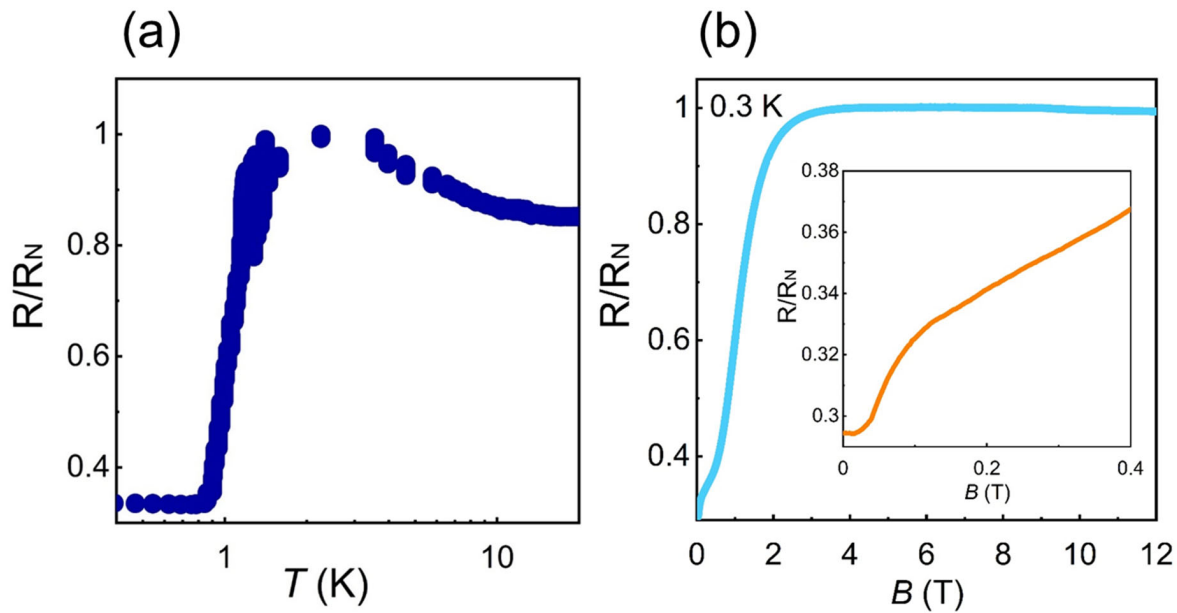

**Figure 2. Superconducting transition at the TiO$_x$/KTaO$_3$(111) interfaces.** (a) Superconducting transition with temperature (mid-point $T_C$ ~1.1 K). A remnant, saturating resistance is observed below the superconducting transition temperature. (b) Longitudinal magnetoresistance shows the superconducting transition and low field signatures of weak anti-localization (inset) at 0.3 K.

The angle-dependent longitudinal magnetoresistance was measured to confirm the presence of the weak anti-localization effect. Figure 3(a) shows a transition from linear positive magnetoresistance to a parabolic behavior, with rotating the magnetic field from out-of-plane to in-plane, respectively, suggesting a 2D weak anti-localization correction. To parse out the superconducting and weak anti-localization components, the longitudinal magnetoresistance was measured and compared between superconducting, KTaO3 (111), and non-superconducting, KTaO3 (100), interfaces, Fig. 3(b). Both interfaces show a positive and linear magnetoresistance with out-of-plane magnetic field. The superconducting interface, however, shows roughly one order of magnitude larger weak anti-localization correction, revealing a potentially enhanced coherence length. The large magnetoresistance at (111) interfaces could also be explained by the pre-formed Cooper pairs. The in-plane magnetoresistance of the superconducting interface shows only 1% positive magnetoresistance at 3 K and 5 T, suggesting that the pair breaking correction could not explain the large positive magnetoresistance at the superconducting interfaces.

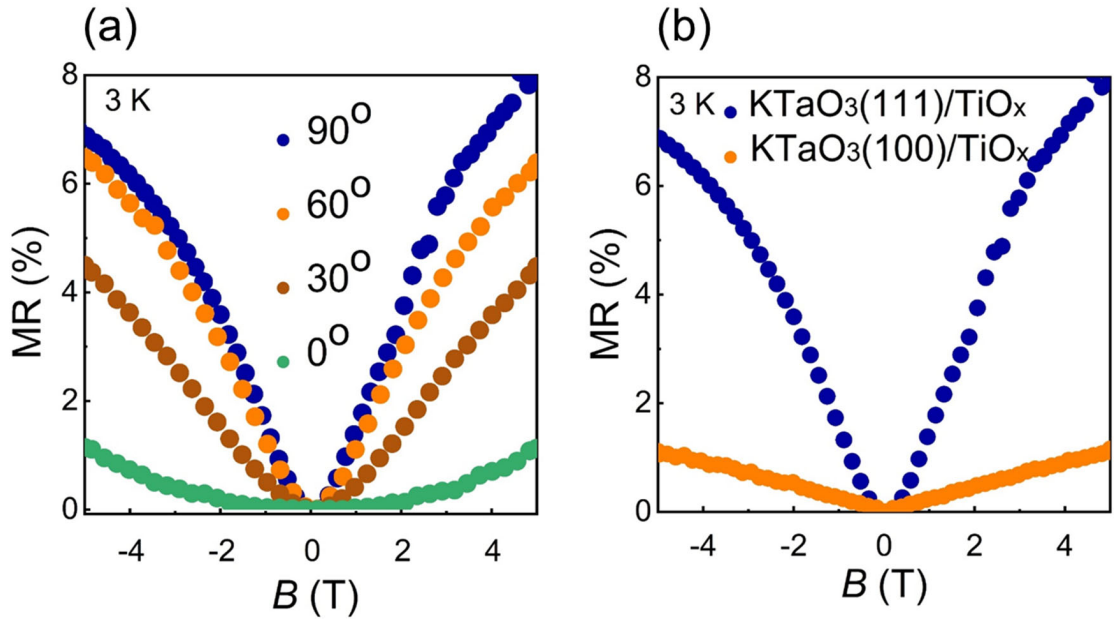

**Figure 3. Weak anti-localization at the TiO$_x$/KTaO$_3$ interfaces.** (a) Angle-dependent magnetoresistance at the TiO$_x$/KTaO$_3$(111) interface. The transition from positive linear (out-of-plane field) to parabolic (in-plane field) suggests a 2D weak anti-localization. (b) Weak anti-localization in superconducting, (111), and non-superconducting, (100), TiO$_x$/KTaO$_3$ interfaces.

**Discussion**

To briefly summarize the results, our main findings are as follows: (i) TiO$_x$/KTaO$_3$(111) interfaces show an abrupt superconducting transition; (ii) The superconducting transition is sensitive to the normal state resistance and a nonzero, saturating resistance persists below the transition temperature; and (iii) Signatures of weak anti-localization below the superconducting transition temperature suggest the coexistence of superconductivity and weak anti-localization.

The first important conclusion from these results is that the emergence of superconductivity at the KTaO3 interfaces depends strongly on the interfacial structure. KTaO3, unlike SrTiO3, does not experience structural instability and remains cubic at low temperature [19,37]. This excludes structural domains as the source of the observed granular superconductivity. Here, the transition could be sensitive to the relaxation time of charge carriers as the interfaces with sheet resistance above the Mott-Ioffe-Regel limit ($h/\tau \sim E_F$, where $h$, $\tau$, and $E_F$ are Planck's constant, relaxation rate, and Fermi energy, respectively) do not show a superconducting transition. This is consistent with a recent report demonstrating electric field control of a superconductor-insulator transition at the LaAlO3/KTaO3(111) interface [23]. Alternatively, the inhomogeneity of $TiO_x$ layer could create an inhomogeneous 2D electron system and superconductivity. The observation of a remnant resistance below the transition temperature means that the superconducting puddles form but fail to coalesce or reach a global phase coherence mediated by Josephson coupling [32,38–40]. Here, the fluctuations of superconducting order parameter in different puddles could limit the long-range phase coupling [32].

Next, we discuss the observation of weak anti-localization and its coexistence with the superconducting phase. 2D electron systems at the KTaO3 interfaces show signatures of weak anti-localization [18,28,29]. Furthermore, topologically nontrivial states are predicted at the KTaO3 (111) surface [11]. The observed weak anti-localization correction, however, is present in both (111) and (100) interfaces. The large weak anti-localization, i.e. coherence length, at the (111) interface could be due to the topologically nontrivial states [9]. Resolving the topological nature of surface electronic states, however, requires further study. The 2D Hikami-Larkin-Nagaoka model [36] does not describe the low-field magneto-conduction behavior at 30 mK, due to the mixed superconducting and weak anti-localization corrections (Supplementary materials, S3). The KTaO3 samples are air-sensitive, and exposure to ambient oxygen fills the surface vacancies, and the 2DEG carrier density declines with exposure to ambient or oxygen annealing [18]. We observe a similar carrier density drop and suppression of the superconductivity in samples without a capping layer due to the strong dependence of superconductivity to carrier density (Supplementary materials, S4). Interestingly, these samples show a linear positive magnetoresistance, after the demise of superconductivity, which could be explained by a 2D Hikami-Larkin-Nagaoka fit, with a resolved coherence length of 103 nm (Supplementary materials, S5), consistent with a previous report [29]. These results confirm the coexistence of weak anti-localization and superconductivity below the superconducting transition temperature at the KTaO$_3$ (111) surfaces.

In summary, our results, especially the coexistence of superconductivity and weak anti-localization, should be of interest for the experimental realization of non-abelian excitations in a single material. We stress that our findings warrant further study of the topological nature of surface states in KTaO$_3$ (111) and the coexistence of topologically nontrivial states with superconductivity. Finally, it would be interesting to explore whether the observed superconductivity is topological.

Acknowledgements

K.A. thanks Anand Bhattacharya for a fruitful discussion. NC team was supported by the U.S. National Science Foundation under Grant No. NSF DMR-1751455.

# Supplementary information

# Coexistence of superconductivity and weak anti-localization at KTaO$_3$ (111) interfaces

Athby H. Al-Tawhid[1], Jesse Kanter[2], Mehdi Hatefipour[2], Divine P. Kumah[3], Javad Shabani[2], and Kaveh Ahadi[1,3,*]

[1]Department of Materials Science and Engineering, North Carolina State University, Raleigh, NC 27695, USA
[2]Center for Quantum Phenomena, Department of Physics, New York University, New York 10003, USA
[3]Department of Physics, North Carolina State University, Raleigh, North Carolina 27695, USA

**Reflection High Energy Electron Diffraction (RHEED)**

Thin films (~3 nm) of amorphous TiO$_x$ were grown on the KTaO$_3$ (111) substrates using an oxide molecular beam epitaxy system. The chamber's base pressure was $2\times10^{-10}$ Torr. The bare substrate shows a clear, streaky RHEED pattern, Fig. S1(a), which disappears gradually during deposition of the TiO$_x$ film, Fig. S1(b), indicating an amorphous TiO$_x$ film growth. Ti acts as the capping layer and oxygen getter, pumping the oxygen atoms from the KTaO$_3$ (111) surface. Oxygen vacancies introduce itinerant electrons to the KTaO$_3$ (111), Ta $5d$ derived surface states.

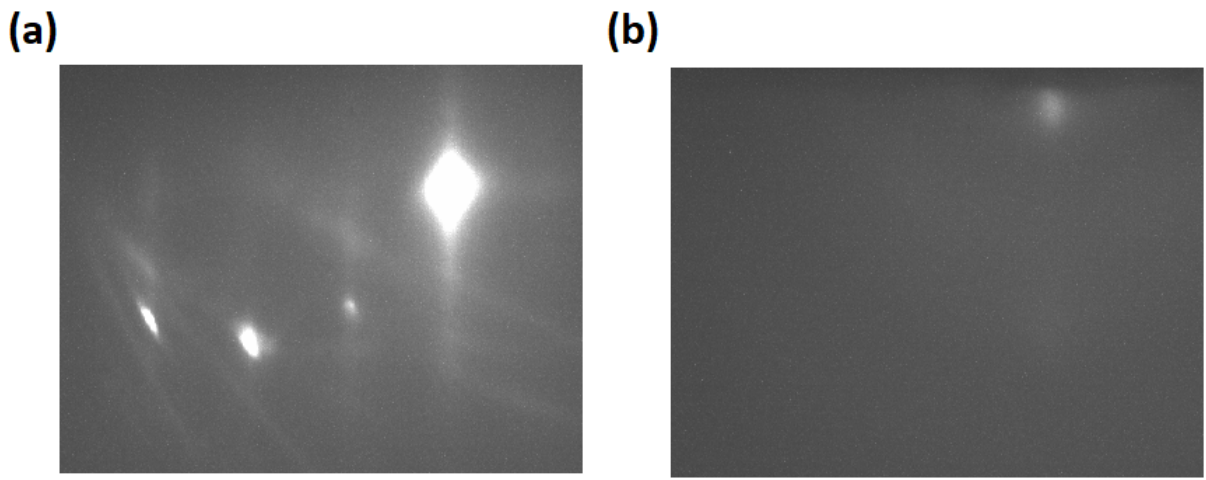

**Figure S1. RHEED patterns during TiO$_x$ growth on KTaO$_3$ (111) substrate.** (a) RHEED pattern of the bare KTaO$_3$ (111) substrate at 400 °C. (b) RHEED image after the deposition of 3 nm TiO$_x$. The disappearance of the RHEED streaks indicates an amorphous TiO$_x$ growth.

## LaCrO$_3$/KTaO$_3$(111) Interfaces

Figure S2 shows the sheet resistance with temperature in LaCrO$_3$/KTaO$_3$(111) heterostructures. The growth details are explained elsewhere [1]. The sheet resistance shows an upturn around 100 K which could be due to the Kondo scattering of the Ta 5$d$ conduction electrons by Cr magnetic moments at the interface. The resistance at 3 K exceeds the 2D Mott-Ioffe-Regel limit. The LaCrO$_3$/KTaO$_3$(111) interfaces do not show a superconducting transition, and the sheet resistance reaches ~10 MΩ/□ at 30 mK. Superconductivity in KTaO$_3$ is highly sensitive to interfacial structure, and similar behavior was recently reported at the LaAlO$_3$/KTaO$_3$(111) interfaces using electric field [2].

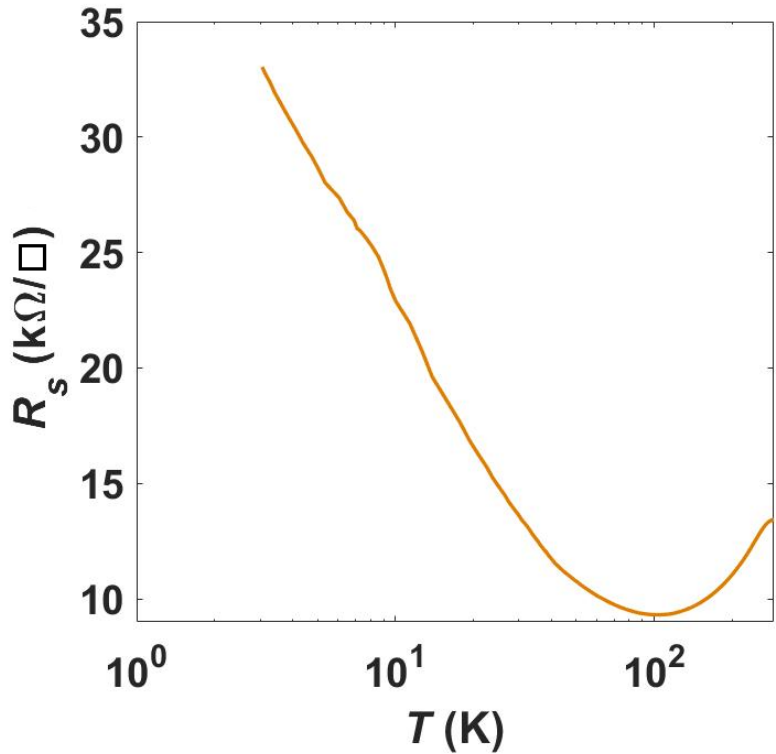

**Figure S2. Temperature-dependent transport at the LaCrO$_3$/KTaO$_3$(111) interface.** Sheet resistance at the LaCrO$_3$/KTaO$_3$(111) interface with temperature. Sample shows a resistance upturn around 100 K.

## Hikami-Larkin-Nagaoka Fit to the Magneto-transport

The Hikami-Larkin-Nagaoka (HLN) model [3] could not explain the magneto-conductance results at 30 mK. We fit the magneto-conductivity at the low-field regime using the 2D HLN model (S1).

$$\Delta\sigma(B) = \frac{\alpha}{\pi}\left[\psi\left(\frac{1}{2} + \frac{\hbar}{4eBl_\varphi}\right) - \ln\left(\frac{\hbar}{4eBl_\varphi}\right)\right] \quad (S1)$$

Here, $\psi$ is the digamma function, $\alpha$ is a constant equal to $-\frac{1}{2}$ and $l_\varphi$ is the coherence length. The least square fit of the 2D HLN model does not capture the low-field behavior. This is due to the mixed weak anti-localization and superconducting corrections to the transport behavior.

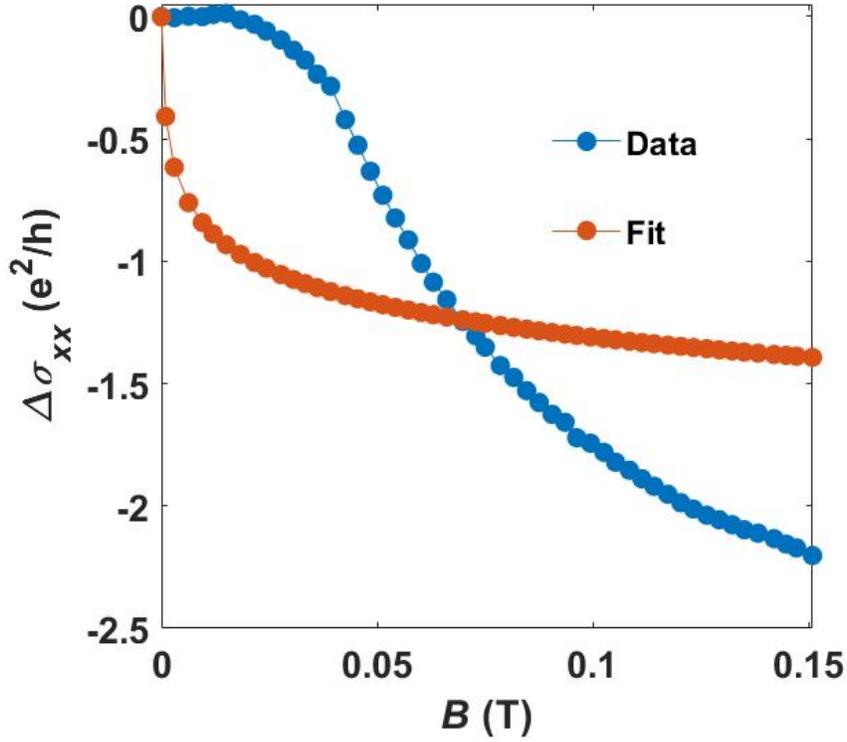

**Figure S3. Magneto-conductivity at the TiO$_x$/KTaO$_3$(111) interfaces with a 2D HLN fit.** The Hikami-Larkin-Nagaoka model could not explain the low-field magneto-conductivity in the failed superconducting state at TiO$_x$/KTaO$_3$(111) interface.

**Effect of Oxygen Vacancy Filling**

We previously showed that the oxygen vacancies at the interface of KTaO$_3$ absorb ambient oxygen, and Hall carrier density decreases, accordingly [1]. The superconducting state is highly sensitive to carrier density, and vanishes in samples without a capping layer [4]. The fresh samples show a failed superconductor state. Superconductivity is suppressed in samples where surface oxygen vacancies are filled. These samples only show linear, positive magnetoresistance.

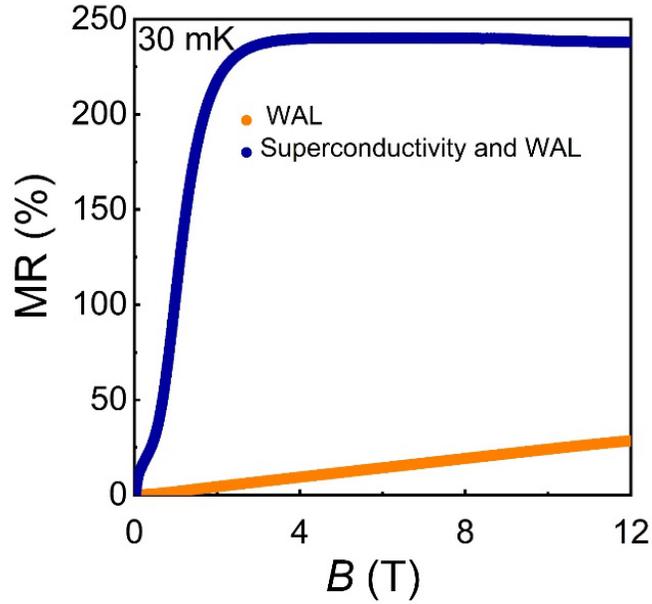

**Figure S4.** Magnetoresistance of TiO$_x$/KTaO$_3$(111) before (blue) and after (orange) oxygen vacancy filling at 30 mK.

The low-field magneto-conductivity at 30 mK was fitted to a 2D HLN model in aged samples. The least square fit to the 2D HLN model (S1), after oxygen vacancy filling, captures the magneto-transport and yields a coherence length of 103 nm, which is consistent with a previous report [5].

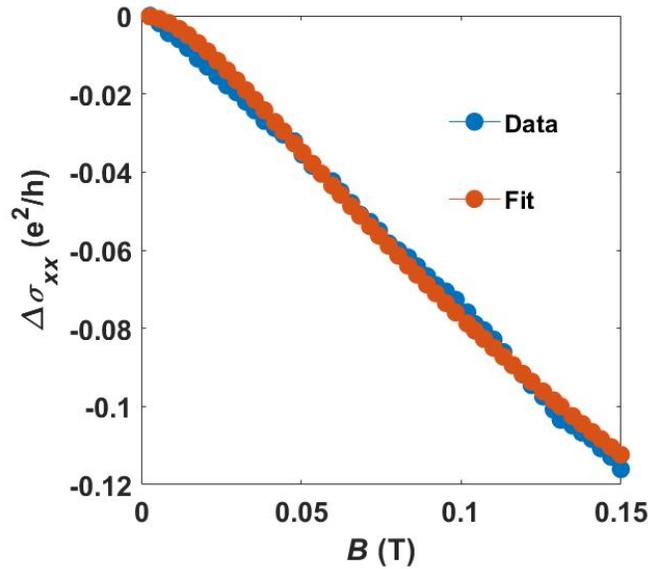

**Figure S5.** Low-field magneto-conductivity of TiO$_x$/KTaO$_3$(111) at 30 mK with a 2D HLN fit, yielding a coherence length of ~103 nm after oxygen vacancy filling.